\documentclass[submission, Phys]{SciPost}

\usepackage[bitstream-charter]{mathdesign}
\hypersetup{
     colorlinks,
     linkcolor={red!50!black},
     citecolor={blue!50!black},
     urlcolor={blue!80!black}
}
\DeclareSymbolFont{usualmathcal}{OMS}{cmsy}{m}{n}
\DeclareSymbolFontAlphabet{\mathcal}{usualmathcal}
\usepackage[font=small]{caption}

\def\vtwo{v_2\{2\}}

\def\vthree{v_3\{2\}}

\def\ptfluct{\delta p_T/\langle p_T\rangle}

\def\tr{\emph{Trajectum}}

\def\mpt{\langle p_T \rangle}

\begin{document}

\begin{flushright}
{\sf CERN-TH-2021-229//MIT-CTP/5383}
\end{flushright}

\begin{center}{\Large \textbf{
Inferring nuclear structure from heavy isobar collisions using \emph{Trajectum}
}}\end{center}

\begin{center}
Govert Nijs\textsuperscript{1$\star$} and
Wilke van der Schee\textsuperscript{2$\dagger$}
\end{center}

\begin{center}
{\bf 1} Center for Theoretical Physics, Massachusetts Institute of Technology, Cambridge, MA 02139, USA
\\
{\bf 2} Theoretical Physics Department, CERN, CH-1211 Gen\`eve 23, Switzerland
\\

${}^\star$ {\small \sf govert@mit.edu}
${}^\dagger$ {\small \sf wilke.van.der.schee@cern.ch}
\end{center}

\begin{center}
\today
\end{center}

\section*{Abstract}
{\bf
Nuclei with equal number of baryons but varying proton number (isobars) have many commonalities, but differ in both electric charge and nuclear structure. Relativistic collisions of such isobars provide unique opportunities to study the variation of the magnetic field, provided the nuclear structure is well understood. In this work we simulate collisions using several state-of-the-art parametrizations of the $^{96}_{40}$Zr and $^{96}_{44}$Ru isobars and show that a comparison with the exciting STAR measurement \cite{2109.00131} of ultrarelativistic collisions can uniquely identify the structure of both isobars. This not only provides an urgently needed understanding of the structure of the Zirconium and Ruthenium isobars, but also paves the way for more detailed studies of nuclear structure using relativistic heavy ion collisions.
}

{\hypersetup{hidelinks}
}

\noindent

\section{Introduction} The structure of nuclei is traditionally studied using low energy experiments \cite{1312.5975,Nayak:2014zba} or global fits of low-energy data using Density Functional Theory (DFT, \cite{Bender:2003jk})\@. Nuclear structure, however, also leaves characteristic imprints on the shape of the quark-gluon plasma (QGP) that is formed shortly after an ultrarelativistic nuclear collision. The resulting anisotropic flow coefficients of the hadrons produced can then in turn be used to infer the shape of this QGP \cite{1301.2826, 1802.04801, 1505.02677}\@.

The recent isobar run of collisions of both $^{96}_{40}$Zr and $^{96}_{44}$Ru at the Relativistic Heavy Ion Collider (RHIC) in New York had a special motivation to search for the chiral magnetic effect (CME, \cite{1511.04050})\@. The idea here is that $_{44}^{96}$Ru has four extra protons and hence generates a 10\% larger magnetic field. All other things equal, the ratio of observables between $_{40}^{96}$Zr and $_{44}^{96}$Ru would provide unique signatures of the magnetic field, whereby comparing like-sign and unlike-sign flow in the direction of the magnetic field can provide convincing evidence for the CME in the presence of a chiral imbalance caused by fluctuations. 

Two crucial features allowed the recent isobar run at RHIC to give results at an unprecedented precision. First of all the two runs were done over the same time span under equal detector settings. While for many measurements systematic uncertainties are often sizeable, the similarity of the two isobars makes that in the ratio almost all these systematics cancel to a high level. Secondly, in \cite{2109.00131} a special effort was made to unambiguously detect the CME by a blinded analysis of $_{40}^{96}$Zr and $_{44}^{96}$Ru collisions, where even the experimentalists did not know which dataset corresponded to which isobar.

Unblinding the analysis did not provide evidence for the CME, but surprisingly $_{44}^{96}$Ru had a higher particle production than $_{40}^{96}$Zr (by up to 10\%) as well as larger elliptic flow ($v_2$, by up to 3\%), albeit having a smaller triangular flow ($v_3$, by 7\% for central collisions). These relatively large differences indicate a difference in nuclear structure whose observable effects are much larger than those induced by the 10\% higher electric charge of $_{44}^{96}$Ru. This  could have been anticipated from nuclear structure analyses \cite{1607.04697, 1710.03086, 1808.06711, 1901.04378, 2102.08158}, though we note significant differences between studies. It is the purpose of this work to utilise these different studies in the state-of-the-art heavy ion code \emph{Trajectum} and show how recent results from DFT can fully explain the STAR findings (see also \cite{1802.02292, 1910.06170, 2109.01631, 2111.15559} for a similar study using the transport model AMPT for $v_2$ and $v_3$)\@.

The hydrodynamic postdiction of the STAR measurements using isobar collisions paves the way for a more elaborate analysis that can also include a magnetic field and potentially a CME\@. Similar techniques, including the study of extremely ultracentral elliptic flow, can also improve our understanding of other nuclei, such as the characteristically ellipsoidal Uranium nucleus collided at RHIC \cite{1305.0173,1406.7522} or the slightly less deformed Xenon nucleus collided at the LHC \cite{1805.01832}\@. We furthermore provide predictions regarding the mean transverse momentum in isobar collisions (see also \cite{2111.14812})\@.

The aim of this work is hence twofold. First, there is an urgent need to clarify the differences between the Ruthenium and Zirconium isobars, which is essential for progress to establish a baseline prediction which can subsequently be used to identify effects of the magnetic field in heavy ion collisions. Second, the isobars provide the perfect playground to use heavy ion collisions and its QGP formation to probe low energy nuclear structure and hence connect with experiments and theory at lower energy relevant for fields as far away as neutron stars (see e.g.~\cite{2001.11228, 2101.03193})\@.

\begin{table}[ht]
\centering
\begin{tabular}{cccccccc}
\hline
\hline
nucleus & $R_p\,$[fm] & $\sigma_p\,$[fm] & $R_n\,$[fm] & $\sigma_n\,$[fm] & $\beta_2$ & $\beta_3$ & $\sigma_\text{AA}\,$[b] \\
\hline
$_{44}^{96}$Ru(1) & 5.085 & 0.46 & 5.085 & 0.46 & 0.158 & 0 & 4.628 \\
$_{40}^{96}$Zr(1) & 5.02 & 0.46 & 5.02 & 0.46 & 0.08 & 0 & 4.540 \\
\hline
$_{44}^{96}$Ru(2) & 5.085 & 0.46 & 5.085 & 0.46 & 0.053 & 0 & 4.605 \\
$_{40}^{96}$Zr(2) & 5.02 & 0.46 & 5.02 & 0.46 & 0.217 & 0 & 4.579 \\
\hline
$_{44}^{96}$Ru(3) & 5.06 & 0.493 & 5.075 & 0.505 & 0 & 0 & 4.734 \\
$_{40}^{96}$Zr(3) & 4.915 & 0.521 & 5.015 & 0.574 & 0 & 0 & 4.860 \\
\hline
$_{44}^{96}$Ru(4) & 5.053 & 0.48 & 5.073 & 0.49 & 0.16 & 0 & 4.701 \\
$_{40}^{96}$Zr(4) & 4.912 & 0.508 & 5.007 & 0.564 & 0.16 & 0 & 4.829 \\
\hline
$_{44}^{96}$Ru(5) & 5.053 & 0.48 & 5.073 & 0.49 & 0.154 & 0 & 4.699 \\
$_{40}^{96}$Zr(5) & 4.912 & 0.508 & 5.007 & 0.564 & 0.062 & 0.202 & 4.871 \\
\hline
\hline
\end{tabular}
\caption{\label{tab:WSparameters}Woods-Saxon parameters and inelastic nucleus-nucleus cross sections for the five cases used in this work, for both $_{44}^{96}$Ru and $_{40}^{96}$Zr\@. The $p$ and $n$ labels denote the different WS distributions used for protons and neutrons, respectively. The $\beta_n$-values are the same for both protons and neutrons. Values for the cross-sections of ZrZr and RuRu collisions are also given.}
\end{table}

\section{The model}

In this work we use the publicly available \emph{Trajectum} 1.2 framework \cite{2010.15134, 2010.15130, 2110.13153}\footnote{The \emph{Trajectum} code can be found at \url{https://sites.google.com/view/govertnijs/trajectum}.} using the maximum likelihood settings and UrQMD for the hadronic gas phase \cite{nucl-th/9803035,hep-ph/9909407} as in \cite{2110.13153}\@. \emph{Trajectum} uses the T\raisebox{-0.5ex}{R}ENTo model \cite{1605.03954} to compute the initial state, where the positions of the nucleons within the nucleus are given by a Woods-Saxon (WS) distribution,
\begin{equation}
\label{eq:roftheta}
P(r,\theta) \propto \left(1 + \exp\left(\frac{r - R(\theta)}{\sigma}\right)\right)^{-1},
\end{equation}
where $R(\theta) = R \cdot \left(1 + \beta_2Y_2^0(\theta) + \beta_3Y_3^0(\theta)\right)$, with $Y_n^0$ spherical harmonics, and the radius $R$, skin depth $\sigma$, quadrupole $\beta_2$ and octupole deformation $\beta_3$ parameters. We use separate WS distributions for protons and neutrons, which allows us to incorporate effects such as a neutron skin.
As in \cite{2110.13153}, T\raisebox{-0.5ex}{R}ENTo also includes a hard-core repulsion implemented through a minimal inter-nucleon distance $d_\text{min}$\@.

In this work, we have used five different cases for the WS distribution, where for both $_{40}^{96}$Zr and $_{44}^{96}$Ru the parameters are given in Tab.~\ref{tab:WSparameters}\@.
Cases 1 and 2 correspond to cases 1 and 2 from \cite{1607.04697,2109.00131}\@.
Case 1 from \cite{1607.04697} is in turn based on $e$--A scattering experiments \cite{1312.5975,Nayak:2014zba}, and case 2 from \cite{1607.04697} is produced from calculations based on the finite-range liquid-drop macroscopic model \cite{nucl-th/9308022}\@.
Cases 3 and 4 come from the top and bottom rows of Tab.~2 from \cite{2103.05595}, which are based on recent DFT\@.
The DFT calculations produce more information than can be captured by the WS parameterization, so in \cite{2103.05595} fits were performed to approximate the DFT result by a WS distribution.
This has been done assuming $\beta_2 = 0$ (case 3) and $\beta_2 = 0.16$ (case 4)\@.
Case 5 is the same as 4, but with $\beta_2$ and $\beta_3$ taken from \cite{2109.01631} (see also \cite{1312.5975,Kibedi:2002ube}; this $\beta_3$ in particular was motivated by the STAR measurement \cite{2109.00131})\@.
Cases 3 to 5 have fitted protons and neutrons separately, which leads to a significant neutron skin for $_{40}^{96}$Zr\@. It is important that in this case the size of the nucleus does not correspond to its charge radius (as in case 1 and 2), but we note that in our computation we treat protons and neutrons equally (see also Appendix \ref{appendix:A})\@.
In this way, case 5 captures both the neutron skin effect, which leads to $_{44}^{96}$Ru being slightly smaller than $_{40}^{96}$Zr, as well as nuclear deformations motivated by experimental measurements.

\begin{figure}
\center
\includegraphics[width=0.6\columnwidth]{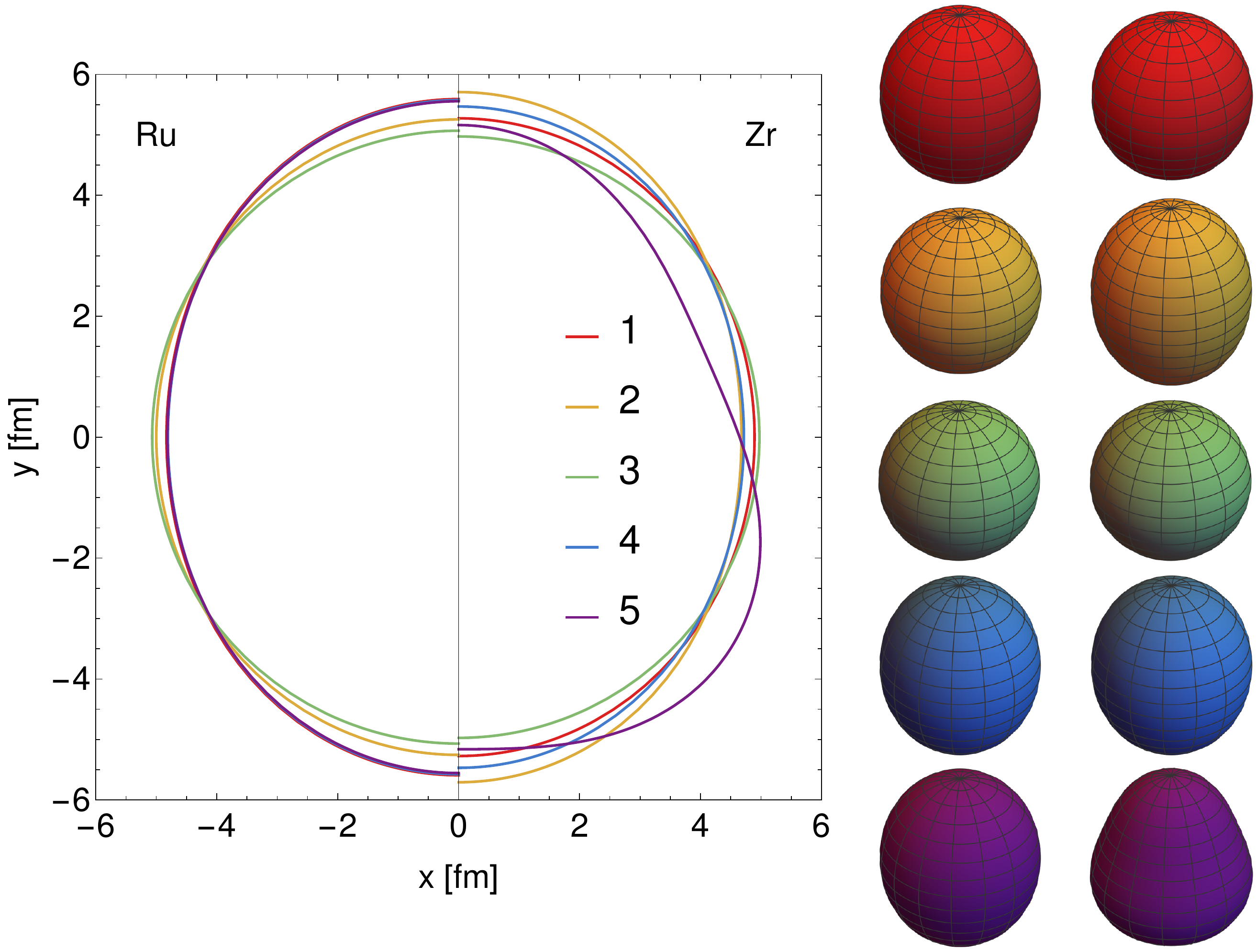}
\caption{\label{fig:isobarnuclei}We show $R(\theta)$ (left) as in \eqref{eq:roftheta} for our 5 different WS distributions of $_{44}^{96}$Ru (left half of plot) and $_{40}^{96}$Zr (right half of plot), where $z = 0$ and nuclei have azimuthal symmetry around the $y$-axis. Here the $R$ parameter has been averaged from $R_p$ and $R_n$ according to the number of protons and neutrons in each nucleus. We also show 3D models (right) of the nuclei.}
\end{figure}

Fig.~\ref{fig:isobarnuclei} illustrates the different sizes and shapes of $_{44}^{96}$Ru (left) and $_{40}^{96}$Zr (right)\@.
Since the WS distribution is separate for protons and neutrons, in Fig.~\ref{fig:isobarnuclei} we use averaged values for the radius $R$, weighted according to the number of protons and neutrons in each nucleus.
For $_{44}^{96}$Ru, it can be seen that cases 1, 4 and 5 are more elliptical than 2 and 3, which is caused by the larger value for $\beta_2$ in cases 1, 4 and 5\@.
For $_{40}^{96}$Zr, in addition to an elliptical shape, one can also see a large triangular deformation for case 5, which is due to the non-zero value for $\beta_3$\@.

Having specified the shape of the isobars we simulate collisions using \tr{}. We stress, however, that we use settings tuned to PbPb collisions at LHC energies, and hence no perfect agreement should be expected at RHIC collision energies. Furthermore, we reduced the normalization of the initial energy density from 18.3 to 7.5, which is loosely tuned to RHIC entropy densities (we find that we underestimate the multiplicity by 18\%). For all cases for both $_{44}^{96}$Ru and $_{40}^{96}$Zr we simulated 0.5M hydro evolutions, except for case 5, where we have 5M hydro events each to be able to probe the 0.1\% most central collisions \footnote{The total computational cost was about 200k CPU hours, \tr{} output files can be found at \href{http://wilkevanderschee.nl/public-codes}{http://wilkevanderschee.nl/public-codes}.}.

\section{Multiplicity and STAR centrality classes}

One of the potentially surprising features of the STAR \cite{2109.00131} result is that the multiplicity in Ruthenium collisions is higher than in Zirconium. This can potentially be explained by Ruthenium's smaller size and hence smaller cross section (see Tab.~\ref{tab:WSparameters})\@.
To compare the exact multiplicities to theoretical computations is however subtle due to various reasons.

\begin{figure}
\center
\includegraphics[width=0.5\textwidth]{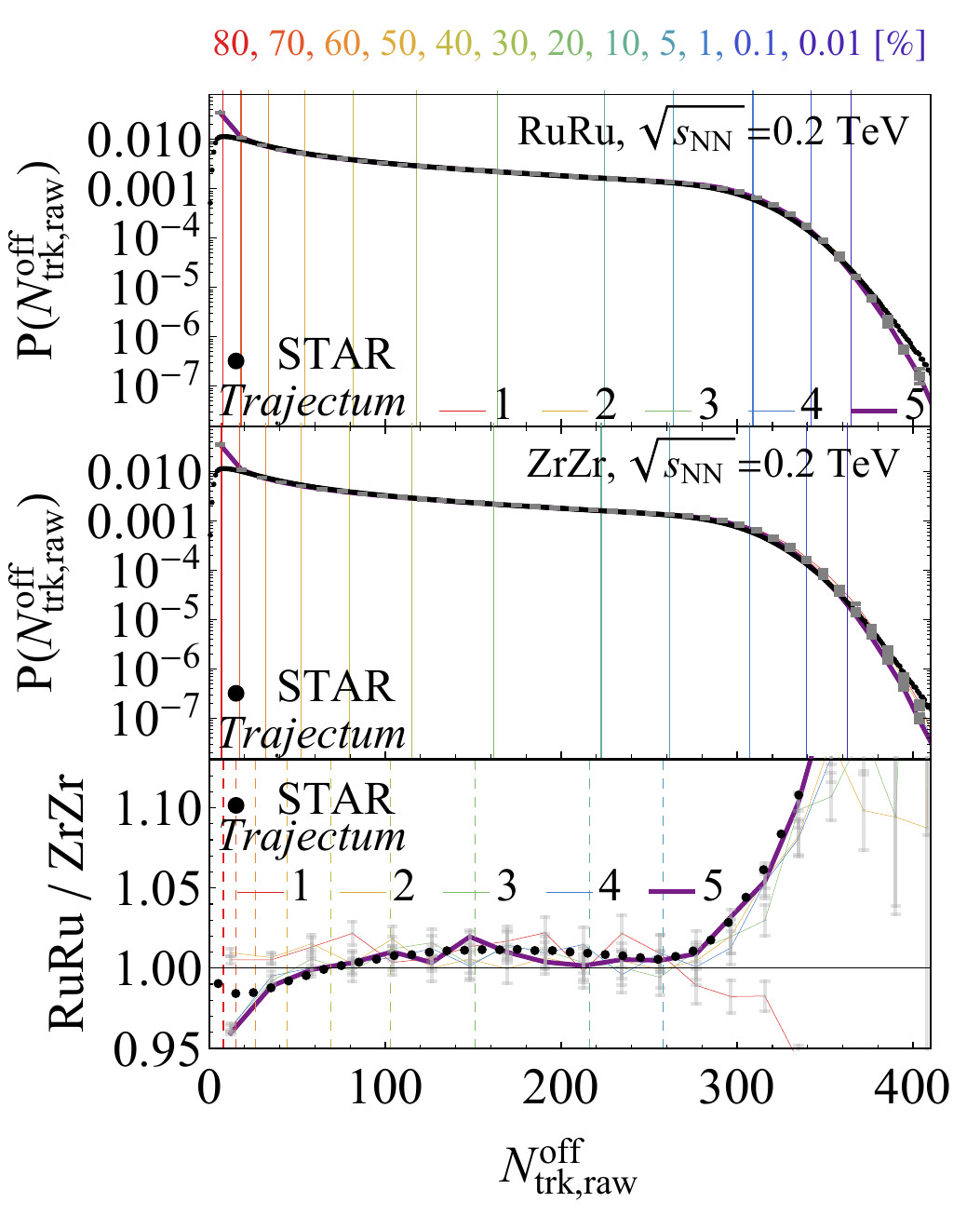}
\caption{\label{fig:v0distribution}We show the STAR $N_{\rm trk,raw}^{\rm off}$ distributions \cite{2109.00131} together with \tr{} results as in Fig.~\ref{fig:v2v3meanpt} together with centrality classes computed by \tr{} (case 5, colored lines) and STAR (dashed)\@. Several corrections have been applied to the theoretical curves, including a correction to the multiplicity (1.21) due to insufficient tuning to RHIC energies, a normalization correction (1.31) to account for STAR not detecting every collision, a correction for detector acceptance (following \cite{0808.2041}) and in the ratio a correction to the experimentally chosen normalizations of $_{44}^{96}$Ru and $_{40}^{96}$Zr in the ratio. After these corrections all cases 3 through 5 have satisfactory agreement.}
\end{figure}

\begin{figure}[t]
\center
\includegraphics[width=0.6\textwidth]{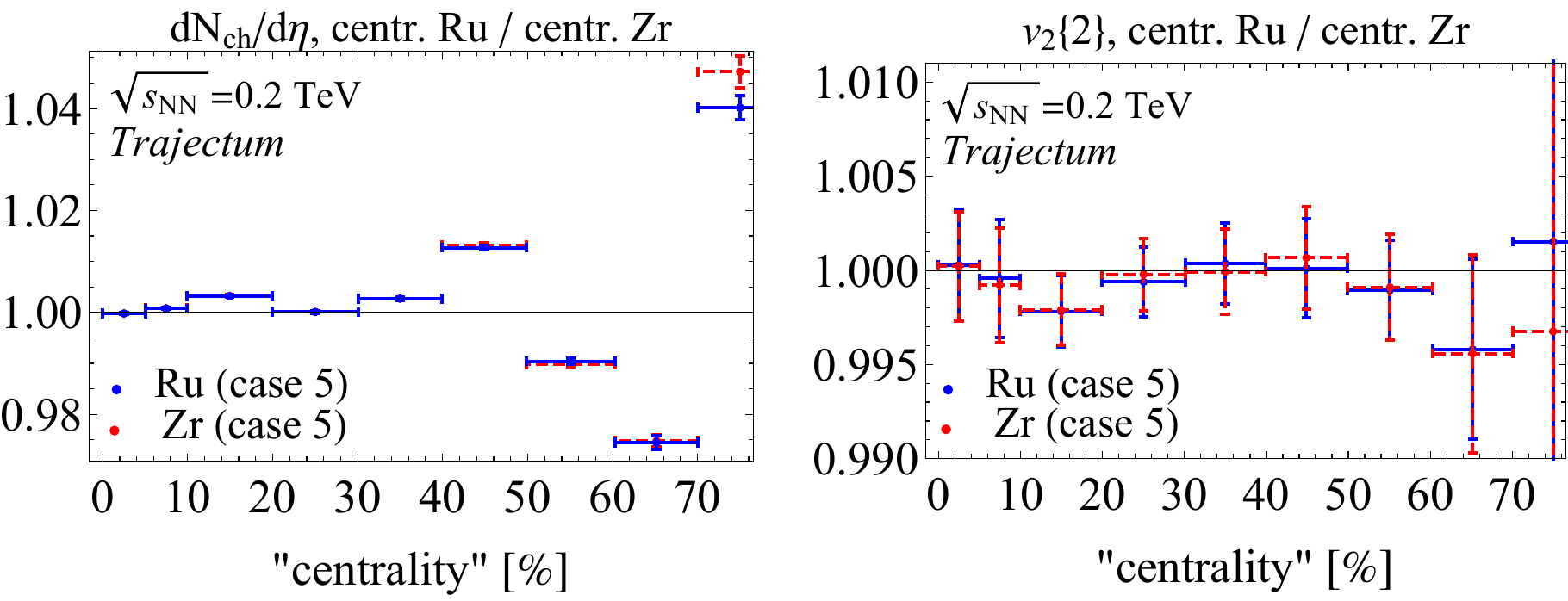}
\caption{\label{fig:starcentrality}We show ratios of multiplicity (left) and elliptic flow (right) using the $_{44}^{96}$Ru and $_{40}^{96}$Zr centrality classes defined in \cite{2109.00131}. Note that we divide $_{44}^{96}$Ru (blue) and $_{40}^{96}$Zr (red) by itself and hence just illustrate the centrality class effect. To emphasize that centrality is in this case a label for a more sophisticated definition we put ``centrality'' in quotation marks on the x-axis. Statistical uncertainties are estimated conservatively by adding up the original fractional uncertainties in quadrature. In reality both quantities use the same event sample and by comparing the Ru and Zr independent samples it can be seen that significant cancellations occur.}
\end{figure}

Ideally an experiment would know the exact luminosity and total hadronic cross section, such that events can be accurately binned into centrality classes. In practice the luminosity is poorly known and at low multiplicity not all events are detected, which is the reason why \tr{} gives a much higher probability for events with fewer than 20 tracks in the detector as seen in Fig.~\ref{fig:v0distribution}\@. The centrality classes in this Figure hence have to be computed using theoretical modelling, which is done here using \tr{} for the theoretical results for both $_{44}^{96}$Ru and $_{40}^{96}$Zr separately and done using an MC Glauber fitting procedure by experiments \cite{2109.00131} (see also \cite{1301.4361})\@. Secondly, the distributions by STAR are raw tracks, which have to be corrected for detector effects (without these corrections it is difficult to fit the multiplicity distribution \cite{2106.02525})\@. Here we apply the inverse correction on \tr{} results using the linear interpolation as in \cite{2109.00131} (original in \cite{0808.2041}) and we apply pseudorapidity $|\eta|<0.5$ and $p_T > 0.2\,$GeV cuts for all charged hadrons. 

After these corrections all in all \tr{} gives an excellent description of the STAR multiplicity in the regime where it is applicable, provided the \tr{} multiplicity is multiplied by 1.21 (this factor arises since the \tr{} norm is only loosely tuned to collisions at RHIC energies) and the normalization of the $N_{\rm trk,raw}^{\rm off}$ distribution changed to 1.31 (this factor captures the fact the STAR curves are normalized and that STAR cannot detect some low multiplicity events).

For the present work we are mostly interested in the isobar ratio. This ratio has an extra subtlety, again related to the normalizations. The experimental curves can only be reliably measured for high multiplicity events and hence a prescription has to be made for the normalization. Here we follow STAR and normalize the models such that the normalization restricted to $N_{\rm trk,raw}^{\rm off}>50$ matches between the model and the data. Since the multiplicity distributions for $_{44}^{96}$Ru and $_{40}^{96}$Zr are rather different this leads to non-trivial correction factors. For the STAR distributions this gives a correction factor of 1.00857 with respect to the  normalized distributions over the full range. For the \tr{} postdictions we simply integrate the $N_{\rm trk,raw}^{\rm off}>50$ distribution and subsequently take the ratio, which leads to correction factors of respectively 
0.99844, 1.0005, 0.97943, 0.9792 and 0.9777\@. We then apply both the experimental and model corrections to the theoretical curves to allow the full comparison in Fig.~\ref{fig:v0distribution} (note that the STAR ratio is simply the ratio of the full normalized raw distributions).

All cases 3 through 5 do a good job reproducing the STAR multiplicity distribution, which illustrates the fact that the multiplicity distribution is mostly determined by the size of the nuclei and not so much by their shape. It is somewhat curious that case 2 performs well in the central region, even though this is not the case for the MC Glauber simulation done in \cite{2109.00131}\@. This could indicate that for the ratio it is necessary to do a full hydrodynamic evolution as opposed to a simple MC Glauber simulation.

Once we identified the multiplicity distribution it is possible to bin our set of events in centrality classes.
For centrality classes that contain low multiplicity events there are essentially two methods to deal with the fact that a centrality definition (involving an integer number of detected particles satisfying some cuts) does not necessarily  align with the chosen centrality classes. The first method, commonly used by experiments, is to compute the observable measured for each integer multiplicity separately. When measuring in a centrality class one would give a particular integer an appropriate weight if it happens to lie on the centrality class boundary.

A second method, employed by \tr{}, is to group events in the centrality classes, whereby events across a centrality class are simply assigned to either class in a properly chosen fraction. Of course for each event such an assignment is arbitrary (in \tr{} it depends on the order of event generation), but since each event is random this arbitrariness does not affect the final observable.

Of course, another option is to conveniently choose centrality classes where the boundaries exactly correspond to the centrality definition chosen. This is for instance done in the STAR paper \cite{2109.00131}, such that it is possible to choose for instance 70.04--79.93\% class for $_{44}^{96}$Ru and 70.00--80.88\% for $_{40}^{96}$Zr\@. In principle there is nothing wrong with such a choice, but when taking a ratio it is important to be careful and remember that one is dividing unequal centrality classes. In \tr{} it is possible to take arbitrary classes, so it is possible to divide such unequal classes, to be able to do a full comparison with the experimental data.

For the multiplicity, being strongly centrality dependent, this is an important consideration, as illustrated in Fig.~\ref{fig:starcentrality} (left)\@. In fact it is the dominant effect going from the 60--70\% class to the 70--80\% class. For elliptic flow ($v_2\{2\}$, shown in Fig.~\ref{fig:starcentrality} (right)) on the other hand the observable is not strongly dependent on centrality and we did not find an effect within our statistical uncertainties.

\section{Flow observables and transverse momentum}

\begin{figure*}[ht]
\center
\includegraphics[width=0.99\textwidth]{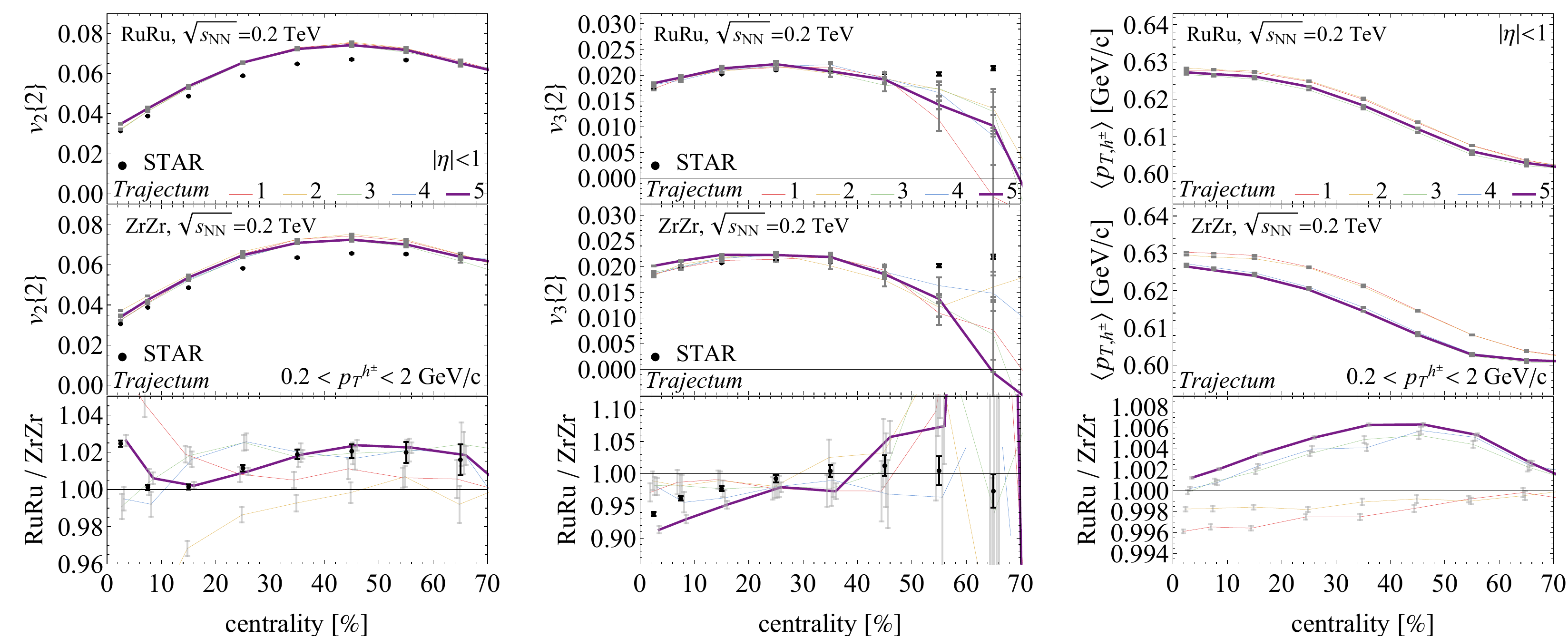}
\caption{\label{fig:v2v3meanpt}We show $\vtwo{}$ (left), $\vthree{}$ (middle) and $\mpt{}$ (right) for $_{44}^{96}$Ru (top), $_{40}^{96}$Zr (middle) and their ratio (bottom) for all five cases of Tab.~\ref{tab:WSparameters} together with STAR data \cite{2109.00131}. Note that \tr{} is only tuned to LHC energies and hence an absolute agreement is not expected. Case 5 is the only case with an octupole deformation $\beta_3$, which leads to a qualitative agreement for the $\vthree{}$ ratios and full consistency for the $\vtwo{}$ ratios. All theoretical uncertainties are statistical only (gray).}
\end{figure*}

Fig.~\ref{fig:v2v3meanpt} shows the elliptic flow $v_2\{2\}$ (left), the triangular flow $v_3\{2\}$ (middle) and the mean transverse momentum $\langle p_T \rangle$ (right) of charged hadrons for both $_{44}^{96}$Ru and $_{40}^{96}$Zr for all five cases and compared to STAR data where available \cite{2109.00131}. For the absolute values we show charged particle $v_2\{2\}$, while for the ratio RuRu/ZrZr we include all particles to increase statistics (charged particle $v_2\{2\}$ is about 1.5\% smaller than the inclusive $v_2\{2\}$, but this effect cancels in the ratio)\@. The absolute postdictions of $v_2\{2\}$ are slightly too high, but we stress again that \tr{} has only been tuned to LHC energies.
More important is the perfect agreement of case 5 for the RuRu/ZrZr ratio of $v_2\{2\}$, capturing both the increase in very central collisions (0--5\%), as well as the increase towards peripheral collisions. The trend for the central $v_3\{2\}$ ratio is also reproduced, but not quantitatively. We note here that case 5 is the only case that includes a $\beta_3$ deformation, which was fitted to these STAR data based on AMPT simulations \cite{2109.01631}\@. We see that the $\beta_3$ deformation is indeed essential to explain the central $v_3\{2\}$, but likely $\beta_3=0.202$ is an overestimate of the deformation (see also Fig.~\ref{fig:isobarnuclei} and see Appendix \ref{appendix:B} for a more quantitative analysis of $\beta_3$).

In the right panel we show the mean transverse momentum, for which STAR data is not yet available (see however \cite{2111.14812})\@. All cases 3 through 5 have similar ratios, which confirms the intuition that mean transverse momentum is not sensitive to the deformations of the nucleus, but that the difference in neutron skin does matter. The effect of the neutron skin is further studied in Appendix \ref{appendix:A}\@.

\begin{figure*}[ht]
\center
\includegraphics[width=0.99\textwidth]{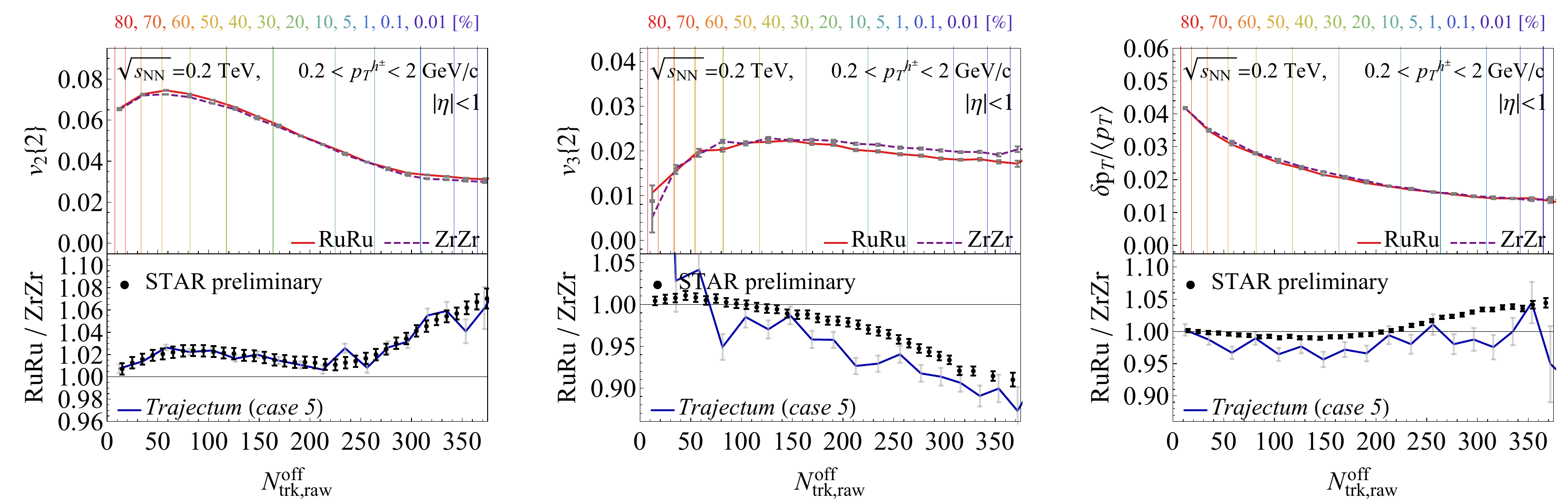}
\caption{\label{fig:ultracentral}For case 5 we show ratios of $\vtwo{}$ (left), $\vthree{}$ (middle) and $\ptfluct{}$ (right) together with preliminary STAR data \cite{STARpreliminary} as a function of $N_{\rm trk,raw}^{\rm off}$\@. This allows to go to extremely ultracentral collisions, up to centrality classes of 0.01\% (see colored lines), which are much more sensitive to deformations in the nuclear shape. Indeed the $\vtwo{}$ and $\vthree{}$ deviate more than in Fig.~\ref{fig:v2v3meanpt}, but nevertheless they are as well described by \tr{} as in Fig.~\ref{fig:v2v3meanpt}. The STAR data contains 0.5\% preliminary systematic uncertainty.
}
\end{figure*}

\section{Ultracentral collisions} As is clear from both Fig.~\ref{fig:isobarnuclei} and \ref{fig:v2v3meanpt} it is interesting to study extremely ultracentral collisions. These have about $N_{\rm trk,raw}^{\rm off}>300$ and correspond to collisions where the two colliding nuclei are completely aligned: there are no spectator nucleons that fly on without interactions. Importantly, this implies that the deformations of the nuclei translate rather precisely into the deformation of the QGP formed, which can then be extracted from the elliptic and triangular flow. Indeed Fig.~\ref{fig:ultracentral} shows a much stronger effect on the ratio of $v_2\{2\}$ and $v_3\{2\}$ as we increase $N_{\rm trk,raw}^{\rm off}$\@. As in Fig.~\ref{fig:v2v3meanpt} the $v_2\{2\}$ dependence is exactly reproduced, whereas $v_3\{2\}$ is overestimated. 

Lastly, we show the fluctuations of the mean transverse momentum in Fig.~\ref{fig:ultracentral} (right) (see also \cite{2110.13153} for similar predictions for PbPb collisions)\@. This observable is statistically difficult, especially to obtain the isobar ratio precisely, but the overall trend seems to be consistent with preliminary STAR data \cite{STARpreliminary}\@.

\begin{figure}
\center
\includegraphics[width=0.6\textwidth]{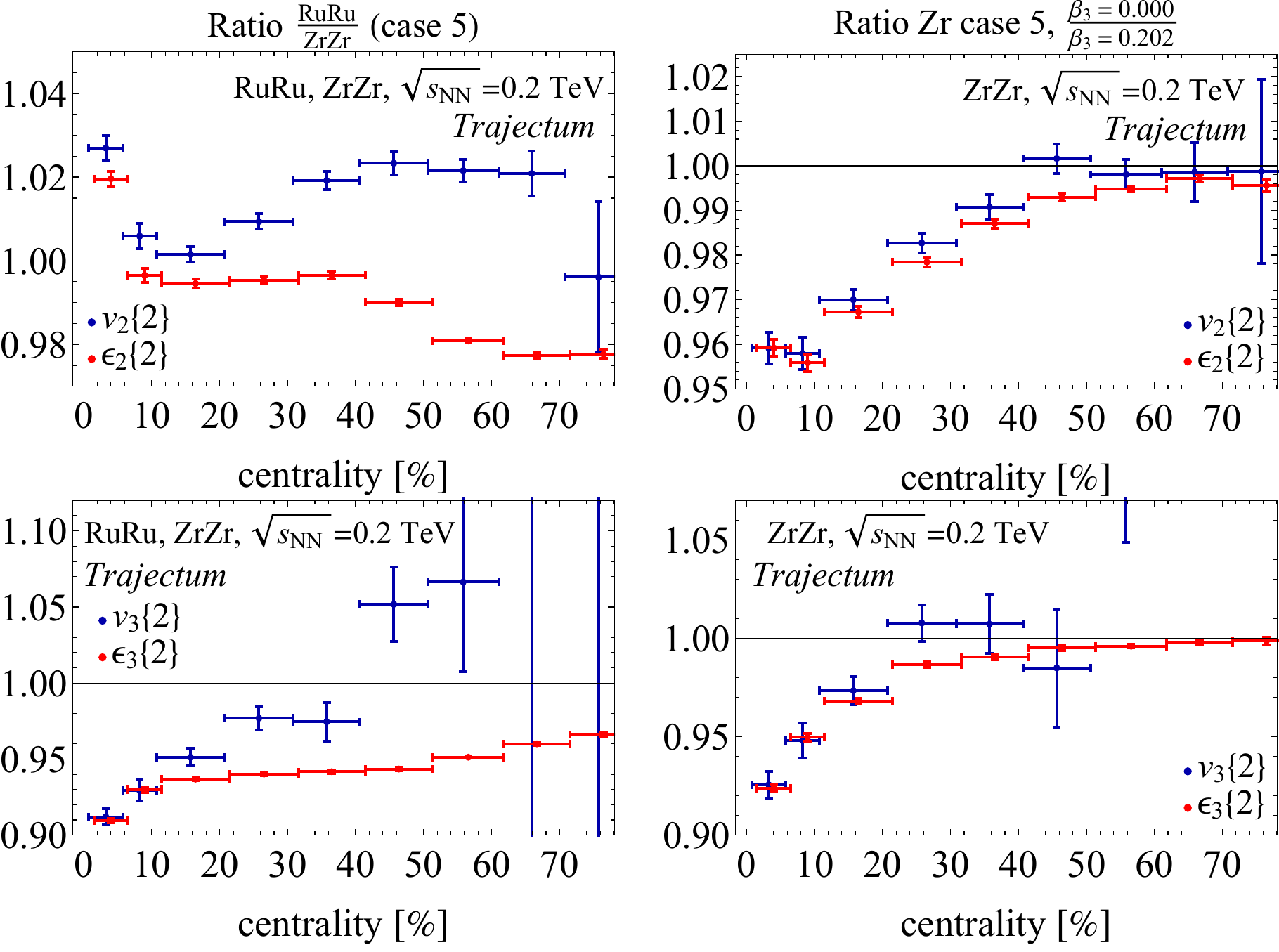}
\caption{\label{fig:vnepsn}It is an interesting question if our $v_2\{2\}$ (top) and $v_3\{2\}$ (bottom) could have been anticipated by initial geometric differences of $\epsilon_n\{n\}$, as in \cite{1505.02677}\@. We show such comparisons for ZrZr/RuRu (case 5, left) and for a Zr-only ratio, where we divide the case of $\beta_3 = 0.202$ with the case $\beta_3 = 0$ (right).}
\end{figure}

\section{Initial eccentricity and hydro response}\label{sec:hydro}

Part of the fit of $\vtwo{}$ and $\vthree{}$ could be anticipated by the changes in the initial geometry of the QGP, as shown in Fig.~\ref{fig:vnepsn}\@. Indeed, in QGP studies the $v_n\{2\}$ distributions usually closely follow the spatial anisotropy distributions $\epsilon_n\{2\}$ so that $v_n\{2\}=\kappa_n \epsilon_n\{2\}$ (see \cite{1505.02677})\@. The coefficients $\kappa_n$ have a strong dependence on the shear viscosity (larger shear viscosity leads to reduced anisotropic flow $v_n$), but a priori it is unclear if $\kappa_n$ depends on the isobar in question. Fig.~\ref{fig:vnepsn} (left) shows indeed that it does, since the $v_n\{2\}$ and $\epsilon_n\{2\}$ isobars are qualitatively different for peripheral collisions (note that for central collisions we find better agreement)\@. This implies that $\kappa_n$ is different for $_{44}^{96}$Ru and $_{40}^{96}$Zr, which likely can be attributed to their different sizes.

This interpretation is further strengthened by  Fig.~\ref{fig:vnepsn} (right), which shows a ratio of the two simulations of $_{40}^{96}$Zr with case 5 and case 5 with $\beta_3=0$ (see also Appendix \ref{appendix:B})\@. In this case the $v_n\{2\}$ and $\epsilon_n\{2\}$ ratios are consistent. Interestingly, the change in $\beta_3$ has a strong effect also on the ellipticity $\epsilon_2\{2\}$\@. We also note that known non-linearities \cite{1505.02677} between $v_n\{2\}$ and $\epsilon_n\{2\}$ do not play a large role here, since indeed $v_n\{2\}$ and $\epsilon_n\{2\}$ change indistinguishably within our statistical uncertainties.

Even though $\kappa_n$ depends on the viscosities we stress that all our isobar ratios do not (strongly) depend on the viscosities, since the viscosities lead to a similar change in $\kappa_n$ for both $_{44}^{96}$Ru and $_{40}^{96}$Zr (see also \cite{2111.14812} for similar observations using AMPT for $\mpt{}$)\@. We show this by explicit computation in Appendix \ref{appendix:B}\@.
More generally, it is expected that the dependence of observables on other model dependencies such as $d_{\rm min}$ in the initial state or other pre-hydrodynamic parameters mostly cancel when taking a ratio of observables from the two isobars.

\section{Discussion}
All of our results agree qualitatively with AMPT studies \cite{2111.15559, 2110.01435} and as such our study is an important confirmation of those conclusions using a true hydrodynamic evolution.
One surprising finding, however, is the non-trivial $\vthree{}$ isobar ratio for central collisions even in the absence of an octupole deformation $\beta_3$\@. This can be seen for all cases in Fig.~\ref{fig:v2v3meanpt}, and also for a high statistics verification in Appendix \ref{appendix:B}\@. It is hence non-trivial to estimate the $\beta_3$ value that would fit the measured $\vthree{}$ ratio. Moreover, as shown in Section \ref{sec:hydro} $\beta_3$ also strongly affects $\vtwo{}$,
which indicates that a full global analysis \cite{Bernhard:2019bmu, 2010.15130,2011.01430} may be needed to obtain even more accurate results. Importantly, a full hydrodynamic analysis seems necessary as just focusing on the $\epsilon_n$ initial anisotropies does not capture the $v_n\{2\}$ dependence quantitatively (see Section \ref{sec:hydro})\@.

In one way the current work provides an accurate analysis of the `baseline' observables for the $_{44}^{96}$Ru and $_{40}^{96}$Zr isobar run performed at RHIC\@. The true motivation for this run was to study the effects of magnetic field, which will only be included in \tr{} in the future. The first step for the isobar run is however to obtain an accurate baseline including the shapes of $_{44}^{96}$Ru and $_{40}^{96}$Zr, which is provided in this work. Importantly, we have seen that this baseline is interesting in its own right, which bridges the fields of relativistic heavy ions and nuclear structure.

\section*{Acknowledgements}
We are very grateful to Chunjian Zhang for providing STAR data and for his explanations. We thank Jiangyong Jia, Rongrong Ma, Rosi Reed, Mike Sas and Prithwish Tribedy for interesting discussions, and Giuliano Giacalone for a careful reading of the manuscript. GN is supported by the U.S. Department of Energy, Office of Science, Office of Nuclear Physics under grant Contract Number DE-SC0011090.

\vspace{2ex} 

\begin{figure*}
\includegraphics[width=0.99\textwidth]{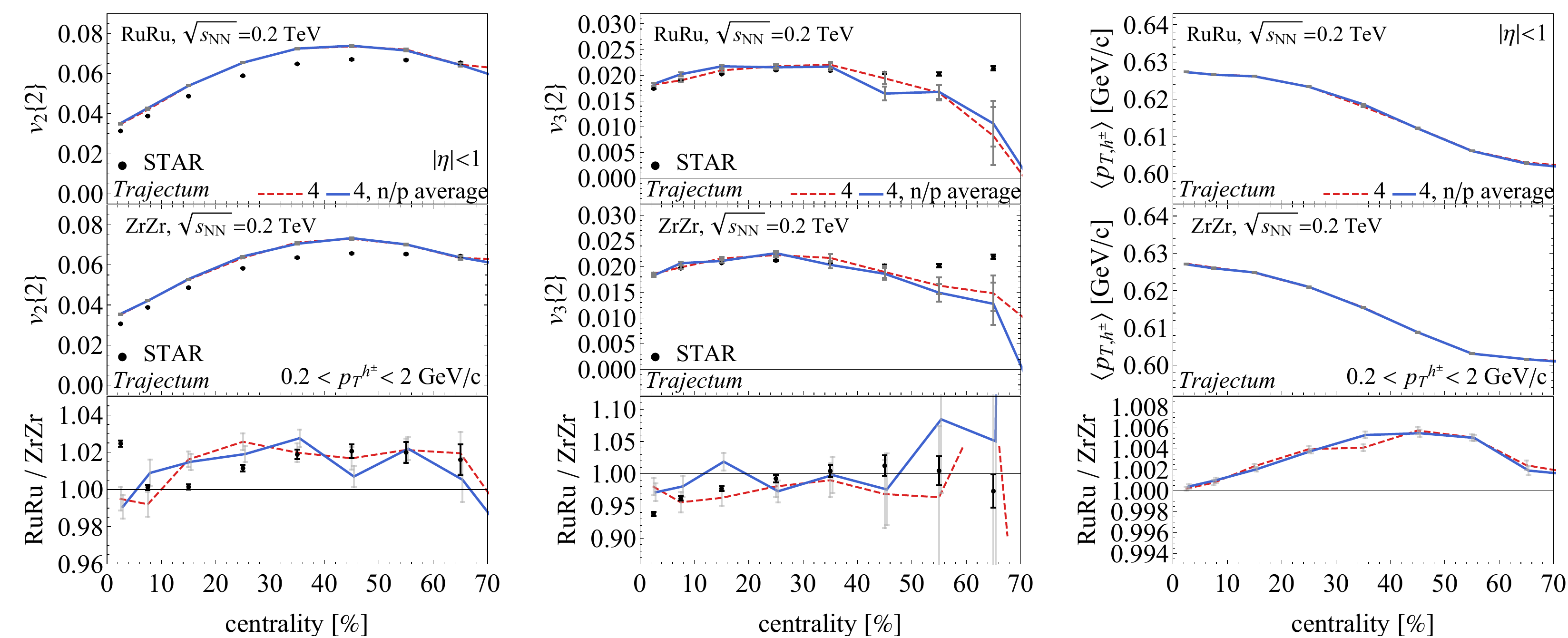}
\caption{\label{fig:neutronskin}We show the effect of using single radius and skin depth parameters in the WS distribution for both neutrons and protons for case 4\@. These are refits to the results from DFT, though they are close to taking the weighted average of the corresponding proton and neutron WS parameters. All results agree within our statistical uncertainties.}
\end{figure*}

\begin{appendix}

\section{Averaging WS proton and neutron values}\label{appendix:A}
The values for WS parameters in Tab.~\ref{tab:WSparameters} treat protons and neutrons separately for cases 3 through 5\@.
Recall that these values were obtained by fitting to a DFT result.
Tab.~2 from \cite{2103.05595} also contains fits which do not treat protons and neutrons separately, but instead describe both using a single WS distribution.
It is important that this single WS distribution still takes the neutron skin into account in the sense that this WS distribution is not fitted to the charge radius, but instead in a sense averaged from the proton and neutron distributions.
This procedure leads to $R = 5.065$, $\sigma = 0.485$ for $_{44}^{96}$Ru and $R = 4.961$, $\sigma = 0.544$ for $_{40}^{96}$Zr\@.
This is not exactly the same as simply taking a weighted average of $R$ and $\sigma$ over the separate proton and neutron WS distributions, which would have given $R = 5.064$, $\sigma = 0.485$ for $_{44}^{96}$Ru and $R = 4.967$, $\sigma = 0.541$ for $_{40}^{96}$Zr, but close enough to justify the term average.
In Fig.~\ref{fig:neutronskin}, we compare the flow and mean transverse momentum for case 4 with separate proton and neutron distributions to the corresponding results with the `averaged' WS distribution.
As the results are compatible with each other within the statistical uncertainties, we conclude that for these nuclei the averaging does not have a big effect on the observables shown, but as is clear from comparing case 3 with case 1 and 2 the neutron skin by itself is important.

\begin{figure*}
\includegraphics[width=0.99\textwidth]{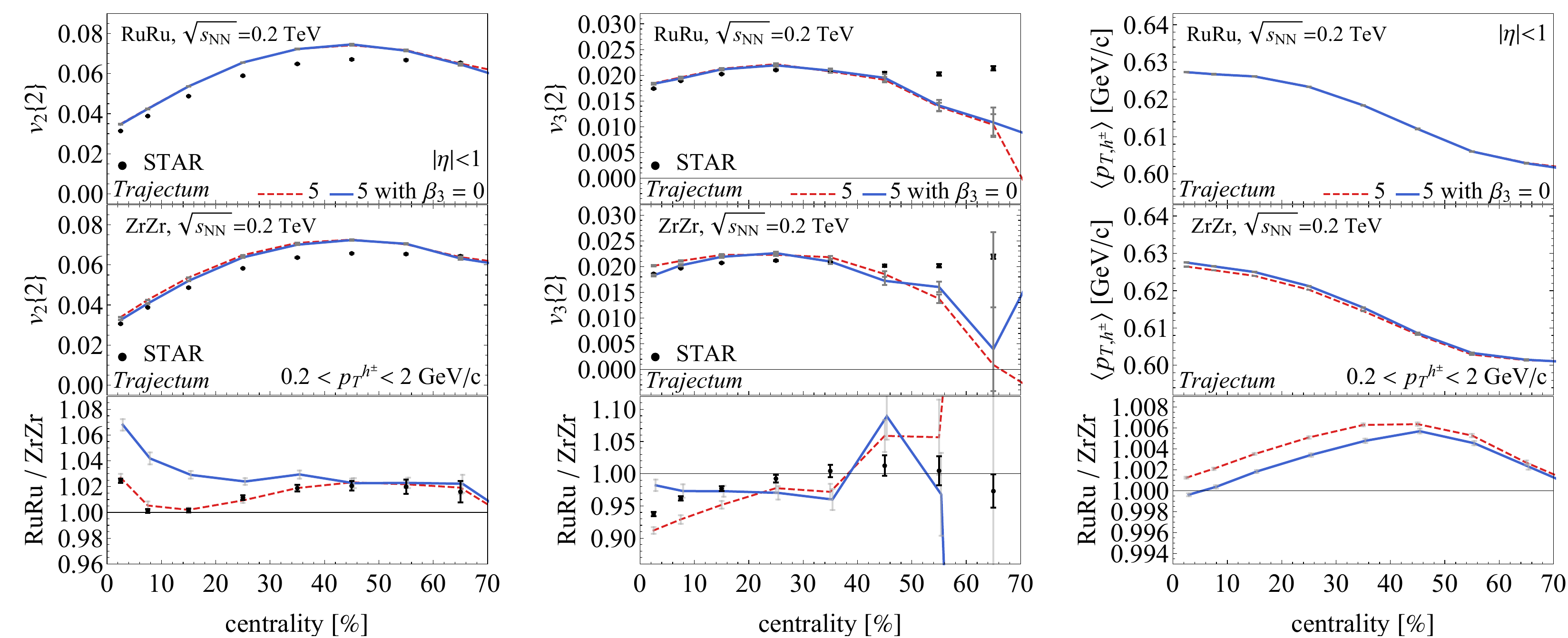}
\caption{\label{fig:beta3effect}We show the effect of turning off $\beta_3$ on $v_2\{2\}$ (left), $v_3\{2\}$ (middle) and hadron $\langle p_T\rangle$ (right) for case 5\@. As expected $\beta_3$ has a strong effect on $\vthree{}$, although it can be seen  that even without $\beta_3$ there is a nontrivial baseline $v_3\{2\}$ ratio (estimated as 0.98 by averaging case 1 through 4 in Fig.~\ref{fig:v2v3meanpt})\@. Importantly, we see that $\beta_3$ has an important effect also on $\vtwo{}$\@.}
\end{figure*}

\begin{figure*}
\includegraphics[width=0.99\textwidth]{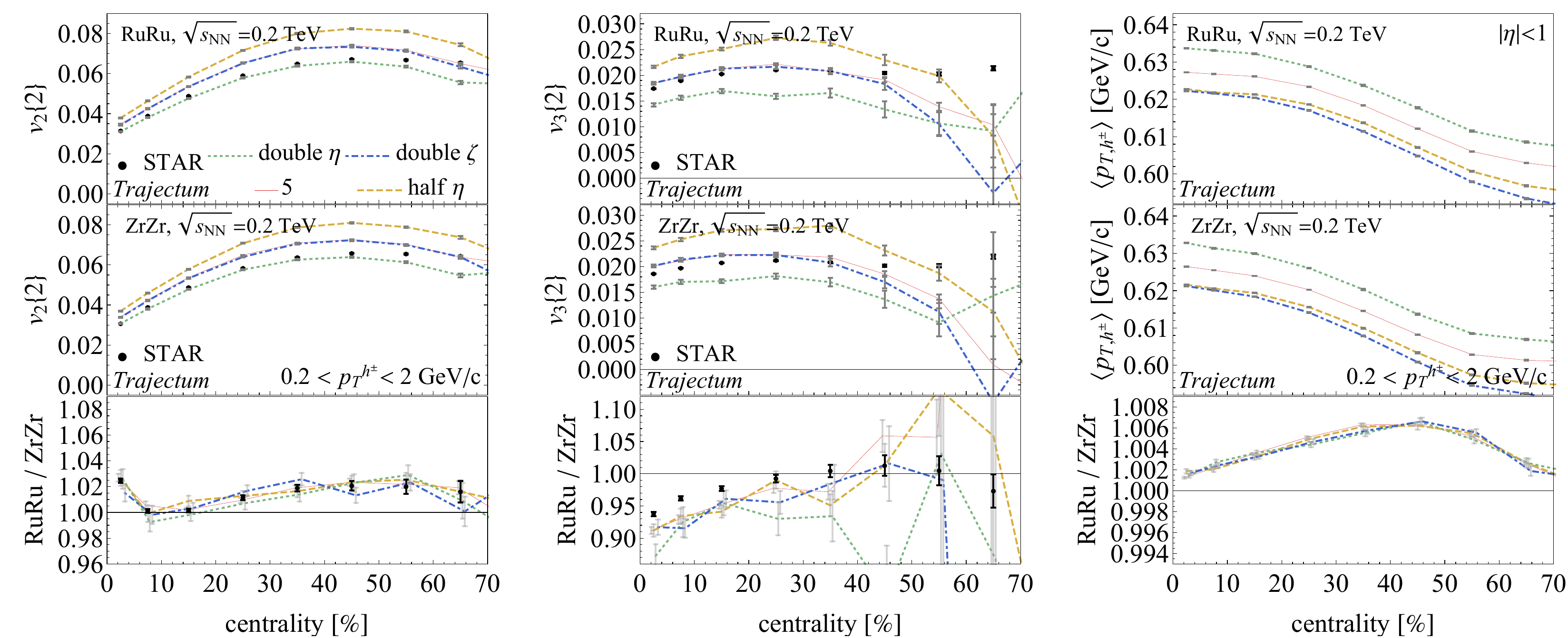}
\caption{\label{fig:etaeffect}We show the effect on $\vtwo{}$ (left), $\vthree{}$ (middle) and $\mpt{}$ of halving and doubling the shear viscosity (dashed and dotted) as well as doubling the bulk viscosity (dashed-dotted)\@. As expected the viscosities have significant effects on all observables, but it can be seen that in the isobar ratio these effects cancel.}
\end{figure*}

\section{Hydrodynamic response to $\beta_3$ and the viscosities.}\label{appendix:B}
Given that case 5 does a much better job than case 4 at describing the $v_2\{2\}$ ratio between RuRu and ZrZr, an interesting question is to what extent this is due to the octupole deformation $\beta_3$\@.
Since cases 4 and 5 differ in both $\beta_2$ and $\beta_3$, it is not immediately obvious how the effects from $\beta_2$ and $\beta_3$ can be disentangled.
To study this in some more detail, in Fig.~\ref{fig:beta3effect}, we show $v_2\{2\}$ (left), $v_3\{2\}$ (middle) and charged hadron $\langle p_T\rangle$ (right) for case 5, as well as a modified case 5 where $\beta_3$ has been set to zero (we use simulations with 5M and 2.5M events respectively)\@. Indeed $\beta_3$ has a strong effect on $\vthree{}$ for central collisions, but even with $\beta_3=0$ the isobar ratio is significantly below unity.
Interestingly, even though $\beta_3$ is a triangular deformation, its effect on $v_2\{2\}$ is also sizable, on the order of a couple percent for more central collisions. Given that $\beta_2$ and $\beta_3$ interplay in their influence on $v_2\{2\}$ and $v_3\{2\}$, this indeed shows, as remarked in the discussion, that decreasing $\beta_3$ to improve the fit of $v_3\{2\}$ would also necessitate a refit of $\beta_2$ to ensure that $v_2\{2\}$ remains in agreement with experimental data.

In Fig.~\ref{fig:etaeffect} we show explicitly that the results on the isobar ratio do not depend on the values of the shear and bulk viscosities (see also \cite{2111.15559} for similar observations using AMPT)\@. This implies that while $\kappa_n$ depends on the viscosities this dependence cancels in the ratio. In particular it implies that $\kappa_n$ consists of terms that depend on the size of the nucleus (which differs between the isobars) and the viscosities (which are independent of the nuclei colliding), but does not (sensitively) depend on a combination of the two.

\end{appendix}

\bibliography{main, manual}

\begin{thebibliography}{10}
\providecommand{\url}[1]{\texttt{#1}}
\providecommand{\urlprefix}{URL }
\expandafter\ifx\csname urlstyle\endcsname\relax
  \providecommand{\doi}[1]{doi:\discretionary{}{}{}#1}\else
  \providecommand{\doi}{doi:\discretionary{}{}{}\begingroup
  \urlstyle{rm}\Url}\fi
\providecommand{\eprint}[2][]{\url{#2}}

\bibitem{2109.00131}
M.~Abdallah \emph{et~al.},
\newblock \emph{{Search for the chiral magnetic effect with isobar collisions
  at $\sqrt {s_{NN}}$=200 GeV by the STAR Collaboration at the BNL Relativistic
  Heavy Ion Collider}},
\newblock Phys. Rev. C \textbf{105}(1), 014901 (2022),
\newblock \doi{10.1103/PhysRevC.105.014901},
\newblock \eprint{2109.00131}.

\bibitem{1312.5975}
B.~Pritychenko, M.~Birch, B.~Singh and M.~Horoi,
\newblock \emph{{Tables of E2 Transition Probabilities from the first $2^{+}$
  States in Even-Even Nuclei}},
\newblock Atom. Data Nucl. Data Tabl. \textbf{107}, 1 (2016),
\newblock \doi{10.1016/j.adt.2015.10.001},
\newblock [Erratum: Atom.Data Nucl.Data Tabl. 114, 371--374 (2017)],
\newblock \eprint{1312.5975}.

\bibitem{Nayak:2014zba}
R.~C. Nayak and S.~Pattnaik,
\newblock \emph{{${\rm B(E2)}\uparrow (0_{1}^{+} \rightarrow 2_{1}^{+})$
  predictions for even\textendash{}even nuclei in the differential equation
  model}},
\newblock Int. J. Mod. Phys. E \textbf{24}(02), 1550011 (2015),
\newblock \doi{10.1142/S0218301315500111},
\newblock \eprint{1405.3191}.

\bibitem{Bender:2003jk}
M.~Bender, P.-H. Heenen and P.-G. Reinhard,
\newblock \emph{{Self-consistent mean-field models for nuclear structure}},
\newblock Rev. Mod. Phys. \textbf{75}, 121 (2003),
\newblock \doi{10.1103/RevModPhys.75.121}.

\bibitem{1301.2826}
U.~Heinz and R.~Snellings,
\newblock \emph{{Collective flow and viscosity in relativistic heavy-ion
  collisions}},
\newblock Ann. Rev. Nucl. Part. Sci. \textbf{63}, 123 (2013),
\newblock \doi{10.1146/annurev-nucl-102212-170540},
\newblock \eprint{1301.2826}.

\bibitem{1802.04801}
W.~Busza, K.~Rajagopal and W.~van~der Schee,
\newblock \emph{{Heavy Ion Collisions: The Big Picture, and the Big
  Questions}},
\newblock Ann. Rev. Nucl. Part. Sci. \textbf{68}, 339 (2018),
\newblock \doi{10.1146/annurev-nucl-101917-020852},
\newblock \eprint{1802.04801}.

\bibitem{1505.02677}
H.~Niemi, K.~J. Eskola and R.~Paatelainen,
\newblock \emph{{Event-by-event fluctuations in a perturbative QCD + saturation
  + hydrodynamics model: Determining QCD matter shear viscosity in
  ultrarelativistic heavy-ion collisions}},
\newblock Phys. Rev. C \textbf{93}(2), 024907 (2016),
\newblock \doi{10.1103/PhysRevC.93.024907},
\newblock \eprint{1505.02677}.

\bibitem{1511.04050}
D.~E. Kharzeev, J.~Liao, S.~A. Voloshin and G.~Wang,
\newblock \emph{{Chiral magnetic and vortical effects in high-energy nuclear
  collisions\textemdash{}A status report}},
\newblock Prog. Part. Nucl. Phys. \textbf{88}, 1 (2016),
\newblock \doi{10.1016/j.ppnp.2016.01.001},
\newblock \eprint{1511.04050}.

\bibitem{1607.04697}
W.-T. Deng, X.-G. Huang, G.-L. Ma and G.~Wang,
\newblock \emph{{Test the chiral magnetic effect with isobaric collisions}},
\newblock Phys. Rev. C \textbf{94}, 041901 (2016),
\newblock \doi{10.1103/PhysRevC.94.041901},
\newblock \eprint{1607.04697}.

\bibitem{1710.03086}
H.-J. Xu, X.~Wang, H.~Li, J.~Zhao, Z.-W. Lin, C.~Shen and F.~Wang,
\newblock \emph{{Importance of isobar density distributions on the chiral
  magnetic effect search}},
\newblock Phys. Rev. Lett. \textbf{121}(2), 022301 (2018),
\newblock \doi{10.1103/PhysRevLett.121.022301},
\newblock \eprint{1710.03086}.

\bibitem{1808.06711}
H.~Li, H.-j. Xu, J.~Zhao, Z.-W. Lin, H.~Zhang, X.~Wang, C.~Shen and F.~Wang,
\newblock \emph{{Multiphase transport model predictions of isobaric collisions
  with nuclear structure from density functional theory}},
\newblock Phys. Rev. C \textbf{98}(5), 054907 (2018),
\newblock \doi{10.1103/PhysRevC.98.054907},
\newblock \eprint{1808.06711}.

\bibitem{1901.04378}
B.~Schenke, C.~Shen and P.~Tribedy,
\newblock \emph{{Multi-particle and charge-dependent azimuthal correlations in
  heavy-ion collisions at the Relativistic Heavy-Ion Collider}},
\newblock Phys. Rev. C \textbf{99}(4), 044908 (2019),
\newblock \doi{10.1103/PhysRevC.99.044908},
\newblock \eprint{1901.04378}.

\bibitem{2102.08158}
G.~Giacalone, J.~Jia and V.~Som\`a,
\newblock \emph{{Accessing the shape of atomic nuclei with relativistic
  collisions of isobars}},
\newblock Phys. Rev. C \textbf{104}(4), L041903 (2021),
\newblock \doi{10.1103/PhysRevC.104.L041903},
\newblock \eprint{2102.08158}.

\bibitem{1802.02292}
W.-T. Deng, X.-G. Huang, G.-L. Ma and G.~Wang,
\newblock \emph{{Predictions for isobaric collisions at $\sqrt{s_{_{\rm NN}}}$
  = 200 GeV from a multiphase transport model}},
\newblock Phys. Rev. C \textbf{97}(4), 044901 (2018),
\newblock \doi{10.1103/PhysRevC.97.044901},
\newblock \eprint{1802.02292}.

\bibitem{1910.06170}
H.~Li, H.-j. Xu, Y.~Zhou, X.~Wang, J.~Zhao, L.-W. Chen and F.~Wang,
\newblock \emph{{Probing the neutron skin with ultrarelativistic isobaric
  collisions}},
\newblock Phys. Rev. Lett. \textbf{125}(22), 222301 (2020),
\newblock \doi{10.1103/PhysRevLett.125.222301},
\newblock \eprint{1910.06170}.

\bibitem{2109.01631}
C.~Zhang and J.~Jia,
\newblock \emph{{Evidence of Quadrupole and Octupole Deformations in Zr96+Zr96
  and Ru96+Ru96 Collisions at Ultrarelativistic Energies}},
\newblock Phys. Rev. Lett. \textbf{128}(2), 022301 (2022),
\newblock \doi{10.1103/PhysRevLett.128.022301},
\newblock \eprint{2109.01631}.

\bibitem{2111.15559}
J.~Jia and C.-J. Zhang,
\newblock \emph{{Scaling approach to nuclear structure in high-energy heavy-ion
  collisions}}  (2021),
\newblock \eprint{2111.15559}.

\bibitem{1305.0173}
Y.~Pandit,
\newblock \emph{{Azimuthal Anisotropy in U+U Collisions at $\sqrt{s_{NN}} = 193
  $ GeV with STAR Detector at RHIC}},
\newblock J. Phys. Conf. Ser. \textbf{458}, 012003 (2013),
\newblock \doi{10.1088/1742-6596/458/1/012003},
\newblock \eprint{1305.0173}.

\bibitem{1406.7522}
H.~Wang and P.~Sorensen,
\newblock \emph{{Azimuthal anisotropy in U+U collisions at STAR}},
\newblock Nucl. Phys. A \textbf{932}, 169 (2014),
\newblock \doi{10.1016/j.nuclphysa.2014.09.111},
\newblock \eprint{1406.7522}.

\bibitem{1805.01832}
S.~Acharya \emph{et~al.},
\newblock \emph{{Anisotropic flow in Xe-Xe collisions at
  $\mathbf{\sqrt{s_{\rm{NN}}} = 5.44}$ TeV}},
\newblock Phys. Lett. B \textbf{784}, 82 (2018),
\newblock \doi{10.1016/j.physletb.2018.06.059},
\newblock \eprint{1805.01832}.

\bibitem{2111.14812}
H.-j. Xu, W.~Zhao, H.~Li, Y.~Zhou, L.-W. Chen and F.~Wang,
\newblock \emph{{Probing nuclear structure with mean transverse momentum in
  relativistic isobar collisions}}  (2021),
\newblock \eprint{2111.14812}.

\bibitem{2001.11228}
M.~Arnould and S.~Goriely,
\newblock \emph{{Astronuclear Physics: a Tale of the Atomic Nuclei in the
  Skies}}  (2020),
\newblock \doi{10.1016/j.ppnp.2020.103766},
\newblock \eprint{2001.11228}.

\bibitem{2101.03193}
B.~T. Reed, F.~J. Fattoyev, C.~J. Horowitz and J.~Piekarewicz,
\newblock \emph{{Implications of PREX-2 on the Equation of State of
  Neutron-Rich Matter}},
\newblock Phys. Rev. Lett. \textbf{126}(17), 172503 (2021),
\newblock \doi{10.1103/PhysRevLett.126.172503},
\newblock \eprint{2101.03193}.

\bibitem{2010.15134}
G.~Nijs, W.~van~der Schee, U.~G\"ursoy and R.~Snellings,
\newblock \emph{{Bayesian analysis of heavy ion collisions with the heavy ion
  computational framework Trajectum}},
\newblock Phys. Rev. C \textbf{103}(5), 054909 (2021),
\newblock \doi{10.1103/PhysRevC.103.054909},
\newblock \eprint{2010.15134}.

\bibitem{2010.15130}
G.~Nijs, W.~van~der Schee, U.~G\"ursoy and R.~Snellings,
\newblock \emph{{Transverse Momentum Differential Global Analysis of Heavy-Ion
  Collisions}},
\newblock Phys. Rev. Lett. \textbf{126}(20), 202301 (2021),
\newblock \doi{10.1103/PhysRevLett.126.202301},
\newblock \eprint{2010.15130}.

\bibitem{2110.13153}
G.~Nijs and W.~van~der Schee,
\newblock \emph{{Predictions and postdictions for relativistic lead and oxygen
  collisions with $Trajectum$}}  (2021),
\newblock \eprint{2110.13153}.

\bibitem{nucl-th/9803035}
S.~A. Bass \emph{et~al.},
\newblock \emph{{Microscopic models for ultrarelativistic heavy ion
  collisions}},
\newblock Prog. Part. Nucl. Phys. \textbf{41}, 255 (1998),
\newblock \doi{10.1016/S0146-6410(98)00058-1},
\newblock \eprint{nucl-th/9803035}.

\bibitem{hep-ph/9909407}
M.~Bleicher \emph{et~al.},
\newblock \emph{{Relativistic hadron hadron collisions in the ultrarelativistic
  quantum molecular dynamics model}},
\newblock J. Phys. G \textbf{25}, 1859 (1999),
\newblock \doi{10.1088/0954-3899/25/9/308},
\newblock \eprint{hep-ph/9909407}.

\bibitem{1605.03954}
J.~E. Bernhard, J.~S. Moreland, S.~A. Bass, J.~Liu and U.~Heinz,
\newblock \emph{{Applying Bayesian parameter estimation to relativistic
  heavy-ion collisions: simultaneous characterization of the initial state and
  quark-gluon plasma medium}},
\newblock Phys. Rev. C \textbf{94}(2), 024907 (2016),
\newblock \doi{10.1103/PhysRevC.94.024907},
\newblock \eprint{1605.03954}.

\bibitem{nucl-th/9308022}
P.~Moller, J.~R. Nix, W.~D. Myers and W.~J. Swiatecki,
\newblock \emph{{Nuclear ground state masses and deformations}},
\newblock Atom. Data Nucl. Data Tabl. \textbf{59}, 185 (1995),
\newblock \doi{10.1006/adnd.1995.1002},
\newblock \eprint{nucl-th/9308022}.

\bibitem{2103.05595}
H.-j. Xu, H.~Li, X.~Wang, C.~Shen and F.~Wang,
\newblock \emph{{Determine the neutron skin type by relativistic isobaric
  collisions}},
\newblock Phys. Lett. B \textbf{819}, 136453 (2021),
\newblock \doi{10.1016/j.physletb.2021.136453},
\newblock \eprint{2103.05595}.

\bibitem{Kibedi:2002ube}
T.~KIB\'EDI and R.~H. SPEAR,
\newblock \emph{{REDUCED ELECTRIC-OCTUPOLE TRANSITION PROBABILITIES, B ( E 3;0
  1 + \textrightarrow{}3 1 \ensuremath{-} )\textemdash{}AN UPDATE}},
\newblock Atom. Data Nucl. Data Tabl. \textbf{80}, 35 (2002),
\newblock \doi{10.1006/adnd.2001.0871}.

\bibitem{0808.2041}
B.~I. Abelev \emph{et~al.},
\newblock \emph{{Systematic Measurements of Identified Particle Spectra in $p
  p, d^+$ Au and Au+Au Collisions from STAR}},
\newblock Phys. Rev. C \textbf{79}, 034909 (2009),
\newblock \doi{10.1103/PhysRevC.79.034909},
\newblock \eprint{0808.2041}.

\bibitem{1301.4361}
B.~Abelev \emph{et~al.},
\newblock \emph{{Centrality determination of Pb-Pb collisions at
  $\sqrt{s_{NN}}$ = 2.76 TeV with ALICE}},
\newblock Phys. Rev. C \textbf{88}(4), 044909 (2013),
\newblock \doi{10.1103/PhysRevC.88.044909},
\newblock \eprint{1301.4361}.

\bibitem{2106.02525}
P.~Carzon, M.~D. Sievert and J.~Noronha-Hostler,
\newblock \emph{{Impact of multiplicity fluctuations on entropy scaling across
  system size}},
\newblock Phys. Rev. C \textbf{105}(1), 014913 (2022),
\newblock \doi{10.1103/PhysRevC.105.014913},
\newblock \eprint{2106.02525}.

\bibitem{STARpreliminary}
\url{https://drupal.star.bnl.gov/STAR/files/ATHIC_Nov_STAR_SBU_ChunjianZhang.pdf}.

\bibitem{2110.01435}
R.~Milton, G.~Wang, M.~Sergeeva, S.~Shi, J.~Liao and H.~Z. Huang,
\newblock \emph{{Utilization of event shape in search of the chiral magnetic
  effect in heavy-ion collisions}},
\newblock Phys. Rev. C \textbf{104}(6), 064906 (2021),
\newblock \doi{10.1103/PhysRevC.104.064906},
\newblock \eprint{2110.01435}.

\bibitem{Bernhard:2019bmu}
J.~E. Bernhard, J.~S. Moreland and S.~A. Bass,
\newblock \emph{{Bayesian estimation of the specific shear and bulk viscosity
  of quark\textendash{}gluon plasma}},
\newblock Nature Phys. \textbf{15}(11), 1113 (2019),
\newblock \doi{10.1038/s41567-019-0611-8}.

\bibitem{2011.01430}
D.~Everett \emph{et~al.},
\newblock \emph{{Multisystem Bayesian constraints on the transport coefficients
  of QCD matter}},
\newblock Phys. Rev. C \textbf{103}(5), 054904 (2021),
\newblock \doi{10.1103/PhysRevC.103.054904},
\newblock \eprint{2011.01430}.

\end{thebibliography}

\end{document}